\documentclass[a4paper,12pt]{article}
\usepackage{amssymb} 
\usepackage{epsfig}
\usepackage{epstopdf}\usepackage{graphicx}
\usepackage{caption}
\usepackage{cite}
\usepackage{subcaption}
\makeatletter
\def\fmslash{\@ifnextchar[{\fmsl@sh}{\fmsl@sh[0mu]}}
\def\fmsl@sh[#1]#2{%
  \mathchoice
    {\@fmsl@sh\displaystyle{#1}{#2}}%
    {\@fmsl@sh\textstyle{#1}{#2}}%
    {\@fmsl@sh\scriptstyle{#1}{#2}}%
    {\@fmsl@sh\scriptscriptstyle{#1}{#2}}}
\def\@fmsl@sh#1#2#3{\m@th\ooalign{$\hfil#1\mkern#2/\hfil$\crcr$#1#3$}}
\makeatother

\usepackage[sumlimits,intlimits,namelimits]{amsmath} 

\usepackage[english]{babel}

\usepackage{graphicx}

\usepackage{geometry}
\numberwithin{equation}{section}
\begin{document}
\begin{titlepage}
\begin{flushright}
SI-HEP-2018-02 \\[0.2cm]
QFET-2018-01 \\[0.2cm]
\today
\end{flushright}

\vspace{1.2cm}
\begin{center}
{\Large\bf 
Reparametrization Invariance and \\[2mm] Partial Re-Summations of the Heavy Quark Expansion}
\end{center}

\vspace{0.5cm}
\begin{center}
{\sc Thomas Mannel }  and {\sc K. Keri Vos }  \\[0.1cm]
{\sf Theoretische Physik 1, Naturwiss. techn. Fakult\"at, \\
Universit\"at Siegen, D-57068 Siegen, Germany}
\end{center}

\vspace{0.8cm}
\begin{abstract}
\vspace{0.2cm}\noindent
We extend existing work on reparametrization invariance (RPI) of the heavy-quark expansion. 
We discuss the total rates of inclusive processes and 
obtain results which have a  manifest RPI  and can be expressed through matrix elements of operators and 
states defined in full QCD. This approach leads to a partial re-summation of higher-order terms in the heavy-quark expansion 
and has the advantage that the number of independent parameters is reduced.  
\end{abstract}

\end{titlepage}

\newpage
\pagenumbering{arabic}
\section{Introduction} 
The Heavy Quark Expansion (HQE) has paved the road to a QCD based calculation of inclusive 
decays of heavy-flavour hadrons and thus has  become an indispensable tool in precision flavour 
physics \cite{Manohar:2000dt}. In particular, the extraction of the CKM parameters from inclusive semileptonic processes 
as well as the search for physics beyond the standard model relies heavily on the HQE, which has 
been pushed to higher orders in the perturbative as well as in the non-perturbative sector to achieve
the highest possible accuracy, e.g.\ in the extraction of $V_{cb}$ \cite{Alberti:2016fba} .  

The HQE can be set up in slightly different ways.  The ``cleanest'' way is to 
extract the complete mass dependence from the matrix elements that define the HQE parameters. 
This is achieved by expressing them in terms of static heavy quark fields defined in Heavy Quark Effective Theory  
(HQET) and hadron states in the infinite mass limit, such that all matrix elements become mass independent. 
However, the
price to be paid is that non-local matrix elements appear which contain the mass dependence 
of the states in full QCD \cite{Bauer:2004ve} . 

This can be avoided by setting up the OPE with fields and states defined in full QCD \cite{Bigi:1994ga}.  
To this end, one removes 
the large part, i.e. the part related to the heavy-quark mass $m$, of the 
heavy quark momentum by a phase factor multiplying the fields and by taking matrix 
elements with states defined in full QCD. In this formulation, the HQE is expressed in terms of local 
matrix elements only; however, these matrix elements still depend on the heavy-quark mass and 
the systematics of the expansion in inverse powers of $m$ are less obvious. In particular, the consideration 
of higher orders requires a careful analysis. Another disadvantage of this approach is that the HQE parameters 
will depend of the quark mass making them non-universal.

Even when the HQE is defined in this second way, we have to introduce a time-like vector $v$ which has to be 
chosen in such a way that the ``residual'' momentum $k = p_Q - m v$ of the heavy quark is a small 
quantity of order $\Lambda_{\rm QCD}$. Although the OPE leading to the HQE is defined as an operator 
relation, the choice of $v$ anticipates that we aim at computing certain matrix elements; in the case of inclusive decays 
these will eventually be the forward matrix elements of hadronic states with the hadron momentum $p_H$;  hence a 
``natural'' choice for the velocity vector is  $v = p_H / m_H$.  

However, the choice of $v$ is not unique; in fact, the final result should not depend on $v$. This Reparametrization Invariance (RPI) \cite{Dugan:1991ak,Chen:1993np} is related to the Lorentz invariance 
of full QCD \cite{Heinonen:2012km}. It has been 
investigated already in  some detail, in particular also for inclusive decays \cite{Luke:1992cs, Man10}.   

A reparametrization (RP), 
i.e. a infinitesimal change in $v$, relates different orders of the HQE. This implies, that a truncation at some order 
$1/m^n$ is invariant under reparametrization only up to terms of order $1/m^{n+1}$.  However, as it has been 
noticed already in \cite{Luke:1992cs, Man10}, RPI fixes the coefficients of towers of operators that are related by reparametrization. 
 
In this paper we show how this can be used to perform a re-summation of these towers of operators in such a way that 
the result can be written in a fully RPI fashion. RPI can be made manifest by expressing the result in terms of matrix elements 
of operators and states of full QCD. This approach also reduces the number of independent parameters, at least for the total rates; 
while in the standard 
formulation this is only implicitly realized, it becomes manifest in our reparametrization-invariant set-up.   

In the next section we give our definition of the reparameterization transformations and apply this in section~\ref{sec:scalar} 
to a scalar toy model of QCD, discussing the total rates. In section~\ref{sec:real} we consider real QCD, i.e including the 
spin of the quarks, again focusing on the total rates.

\section{Reparametrization Transformation}
We start from the equation of motion  for the heavy quark field $Q$, which is the Dirac equation 
\begin{equation} \label{DE} 
(i \fmslash{D} - m) Q(x) = 0 
\end{equation} 
where $D = \partial - i g_s A$ is the QCD covariant derivative with the gluon field $A$. In order to set up the HQE, we introduce a time-like vector $v$, which we use to split the 
momentum $p_Q$ of the heavy quark according to $p_Q = m v +k$. This is achieved by a re-definition of 
the heavy quark field $Q$ according to 
\begin{equation} \label{phaseredef} 
Q(x) = \exp(-i m (v \cdot x)) Q_v (x)   \ ,
\end{equation} 
which implies
\begin{equation} 
(i D_\mu) Q(x) = \exp(-i m (v \cdot x)) (i D_\mu + m v_\mu) Q_v (x)    \, , 
\end{equation} 
corresponding to the decomposition of the heavy-quark momentum with $k \sim iD$.  

Note that the field $Q_v$ is still the field in full QCD; its equations of motion can be derived from (\ref{DE}) 
and read 
\begin{eqnarray}
Q_v &=& \fmslash{v}  Q_v + \frac{i\fmslash{D}}{m} Q_v   \label{eom1} \\ 
(ivD) Q_v &=& - \frac{1}{2m} (i\fmslash{D})   (i\fmslash{D})  Q_v 
= - \frac{1}{2m} (iD)^2 Q_v - \frac{1}{2m} (\sigma \cdot G) Q_v   \label{eom2}
\end{eqnarray} 
where 
\begin{equation} 
 (\sigma \cdot G)  \equiv  (-i \sigma_{\mu \nu}) (iD^\mu)(iD^\nu) \ , \quad\quad \gamma_\mu \gamma_\nu = g_{\mu\nu} + (-i\sigma_{\mu\nu}) \ .
\end{equation} 


The reparametrization transformation  $\delta_{\rm RP}$  corresponding to an infinitesimal change  
$ v_\mu \longrightarrow v_\mu + \delta v_\mu $ is thus
\begin{eqnarray}
&& \delta_{\rm RP} \, v_\mu = \delta v_\mu  \quad  \mbox{with} \quad v \cdot \delta v = 0   \label{RPT1}\\
&& \delta_{\rm RP} \, i D_\mu = - m \delta v_\mu  \label{RPT2}\\ 
&& \delta_{\rm RP} \, Q_v (x) =   i m (x \cdot \delta v) Q_v (x)  \quad \mbox{in particular} \quad \delta_{\rm RP} \,  Q_v (0) = 0 \ .
 \label{RPT3}
\end{eqnarray}   
Note that (\ref{RPT2}) actually follows from from (\ref{RPT3}). 

For the scalar quarks, which we discuss as a toy model, the equation of motion is
 \begin{equation} \label{SG} 
 [(iD)^2-m^2] \phi = 0  \ ,
 \end{equation}  
 where $\phi$ is the field of the scalar quark. 
The redefinition of the field is analogous to (\ref{phaseredef})  
\begin{equation} \label{rescs}
\phi (x) = \frac{1}{\sqrt{2m}} \exp [ -i m (vx)] \phi_v (x) \ ,
\end{equation}
where we have introduced a normalization factor. Note that due to this factor, 
the field $\phi_v$ has a different mass dimension compared to $\phi$:  
${\rm dim}[\phi_v] = 3/2$, while  ${\rm dim}[\phi] = 1$. 
This leads to 
\begin{equation} \label{eoms}
[(iD_\mu + m v_\mu)^2 - m^2 ] \phi_v = 2 m (ivD) \phi_v + (iD)^2 \phi_v = 0 \quad \mbox{or} \quad 
(ivD) \phi_v = - \frac{1}{2m} (iD)^2 \phi_v  \, .
\end{equation} 
The additional factor $\sqrt{2 m}$  in (\ref{rescs}) serves to have the proper normalization of the static 
Lagrangian 
\begin{equation}
{\cal L} =  \phi^\dagger [(iD)^2-m^2] \phi = \phi_v^\dagger (ivD) \phi_v + \cdots  \ .
\end{equation} 

In the scalar case the RP transformation $\delta_{\rm RP}$ remains the same with the obvious replacement 
$Q_v \to \phi_v$.

\section{Toy Model: Scalar Quarks} \label{sec:scalar} 
Before considering full QCD, it is instructive to consider scalar quarks, 
which avoids the unnecessary complications induced by the quark spin. 
To be explicit, we define a decay of such a scalar quark into a lighter scalar quark and a particle without QCD 
interactions, mimicking a semileptonic decay of a real heavy quark. 

\subsection{Reparametrization Invariance} 
We consider the decay of the scalar quark into two lighter scalars $\psi$  and $\ell$ where only one of the two decay 
products  ($\psi$) is a color triplet. Thus we consider an effective Hamiltonian of the form
\begin{equation}
H_{\rm eff} =  g (\phi^\dagger \psi) \ell \ .   
\end{equation} 

With this Hamiltonian we can write total and the differential inclusive rates. We start from the operator 
\begin{equation} \label{intro}
R (q) = \int d^4 x \, e^{iqx} \, T[ (\phi^\dagger (x) \psi (x) ) \, (\psi^\dagger (0) \phi(0) ) ]  \ ,
\end{equation}  
where $q$ is the momentum transfer to the (color-neutral) $\ell$ particle.  Clearly this expression is independent of $v$ 
and thus RPI. 

Inserting the re-scaling (\ref{rescs}) we obtain
\begin{equation} \label{intro1}
R (S) = \int d^4 x \, \frac{1}{2m}e^{-i m (S \cdot x)} \, T[ (\phi^\dagger_v (x) \psi (x) ) \, (\psi^\dagger (0) \phi_v(0) ) ]  \ ,
\end{equation} 
where 
\begin{equation} \label{Sdef} 
S = v - \frac{q}{m}  \ .
\end{equation}
This relation is the starting point of an OPE, leading to a series in inverse powers of $m$. This OPE takes the form  
\begin{equation} \label{opes}
R(S)  =  \sum_{n,i}  C_i^{(n)} (S)  {\cal O}_i^{(n)}   \ , 
\end{equation} 
where the ${\cal O}^{(n)}_i$ are local operators of dimension $n+3$ and the coefficients $C^{(n)}_i$ depend on 
$S$ and are of order $1/m^{n+3}$, assuming dimensionless $R$. At tree level, the operators ${\cal O}^{(n)}_i$
can be written in terms of the fields $\phi_v$ and a chain of covariant derivatives. We get
\begin{equation}\label{opes1}
 R(S)  =    
\sum_{n=0}^\infty  C_{\mu_1 \cdots \mu_n}^{(n)} (S)  \, \,  \phi^\dagger_v (iD^{\mu_1} \cdots iD^{\mu_n})  \phi_v \ ,
\end{equation} 
with $ \phi_v =  \phi_v(0)$. 
Note that the rate is obtained by taking the discontinuity of $R$; since the rate has to be real, the operators appearing in the 
OPE have to be hermitian, so the forward matrix elements are real. This has consequences for the coefficients 
$C_{\mu_1 \cdots \mu_n}^{(n)} (S) $ which we will exploit later on.    

The relation (\ref{opes}) is RPI as long as the full sum is taken into account.
The key observation is that the the RP transformation relates subsequent orders in the $1/m$ expansion. In fact, we have 
\begin{eqnarray}
0 &=& \delta_{\rm RP} R(S) = 
 \sum_{n=0}^\infty  \left[ \delta_{\rm RP} C_{\mu_1 \cdots \mu_n}^{(n)} \right]   \phi^\dagger_v (iD^{\mu_1} \cdots iD^{\mu_n})  \phi_v  
 \\ 
 && \qquad  \qquad \, \, + \nonumber
 \sum_{n=0}^\infty  C_{\mu_1 \cdots \mu_n}^{(n)}   \left[ \delta_{\rm RP} \phi^\dagger_v (iD^{\mu_1} \cdots iD^{\mu_n})  \phi_v \right]
 \\ 
&=&   \sum_{n=0}^\infty  \left[ \delta_{\rm RP} C_{\mu_1 \cdots \mu_n}^{(n)} \right]  \phi^\dagger_v (iD^{\mu_1} \cdots iD^{\mu_n})  \phi_v 
\nonumber \\ \nonumber 
&& - m   \sum_{n=0}^\infty  C_{\mu_1 \cdots \mu_n}^{(n)}    
 \left[ \delta v^{\mu_1} \, \phi^\dagger_v (i D^{\mu_2}) \cdots  (i D^{\mu_n}) \phi_v  
  + \delta v^{\mu_2} \, \phi_v^\dagger (i D^{\mu_1})(i D^{\mu_3}) \cdots  (i D^{\mu_n})  \phi_v   \right. \nonumber \\  
&& \left. \qquad \qquad \qquad  \cdots + \delta v^{\mu_n} \, \phi^\dagger_v (i D^{\mu_1}) \cdots  (i D^{\mu_{n-1}})  \phi_v \right]  \nonumber \ .
\end{eqnarray} 

In order to achieve the cancellation between the different orders in the OPE, the coefficients have to satisfy the relation  
\begin{eqnarray} \label{RPI1}
\delta_{\rm RP} C_{\mu_1 \cdots \mu_n}^{(n)} (S) &=& 
m \, \delta v^{\alpha}  \left( C_{\alpha \mu_1 \cdots \mu_n}^{(n+1)} (S)  +  C_{ \mu_1 \alpha \mu_2 \cdots \mu_n}^{(n+1)} (S)    
+ \cdots + C_{ \mu_1 \cdots \mu_n \alpha}^{(n+1)} (S)  \right)   \, ,   
\end{eqnarray}   
where we have $\delta_{\rm RP} S = \delta v$ from (\ref{Sdef}). 

This is a remarkable relation, relating subsequent orders in the $1/m$ expansion. Furthermore, although it is derived 
using the tree level expression for the operators, its extension to the general case is straightforward.

\subsection{Total Rate} 
Upon integration over the particle $\ell$ we get the OPE for the total rate;
\begin{equation} \label{opes1a}
R   =  \sum_{n=0}^\infty  c_{\mu_1 \cdots \mu_n}^{(n)} (v)  \, \,  \phi^\dagger_v (iD^{\mu_1} \cdots iD^{\mu_n})  \phi_v \ ,
\end{equation} 
where the coefficients $c^{(n)}$ now depend only on $v$. We can decompose the $c^{(n)}$ by writing a linear combination 
with all possible tensor structures composed of $v_\mu$ and $g_{\mu \nu}$, where the coefficients will just be numbers since 
$v^2 = 1$. 

We shall explicitly consider terms up to fourth order, which read   
\begin{eqnarray}
R &=& c^{(0)} \phi_v^\dagger \phi_v + c^{(1)}_\mu  \, \phi^\dagger_v (iD^\mu) \phi_v +  
c^{(2)}_{\mu \nu}  \, \phi^\dagger_v (iD^\mu) (iD^\nu)  \phi_v  \nonumber \\ 
&& + c^{(3)}_{\mu \alpha \nu}  \, \phi^\dagger_v (iD^\mu) (iD^\alpha) (iD^\nu)  \phi_v 
+ c^{(4)}_{\mu \alpha \beta \nu}  \, \phi^\dagger_v (iD^\mu) (iD^\alpha)(iD^\beta) (iD^\nu)  \phi_v 
+ \cdots 
\end{eqnarray} 
and discuss, how this finally generalizes to arbitrary order. 

The tensor decomposition of the $c^{(n)}$, taking into account that only hermitian operators can appear, reads
\begin{eqnarray}  
c^{(0)} (v) &=& a^{(0)} \\  
 c^{(1)}_\mu (v) &=& a^{(1)} v_\mu \\ 
 c^{(2)}_{\mu \nu} (v) &=& a^{(2)} g_{\mu \nu} +  b^{(2)} v_\mu v_\nu  \\ 
 c^{(3)}_{\mu \alpha \nu}  (v) &=& x_1^{(3)}  v_\alpha g_{\mu \nu} + x_2^{(3)} \left[  v_\nu g_{\mu \alpha}  +  v_\mu g_{\nu \alpha} \right] 
 + x_3^{(3)}  v_\mu v_\alpha v_\nu  \label{c3}   \\ 
c^{(4)}_{\mu \alpha \beta \nu} (v)
&=& y^{(4)}_1 g_{\mu \nu} g_{\alpha \beta} +  y^{(4)}_2 g_{\mu \alpha} g_{\nu \beta} + y^{(4)}_3 g_{\mu \beta} g_{\nu \alpha}  
\nonumber \\  \nonumber
&&    + z^{(4)}_1  v_\alpha v_\beta g_{\mu \nu} 
         + z^{(4)}_2 v_\mu v_\nu g_{\alpha \beta}  \nonumber \\ 
&&    + z^{(4)}_3  \left[  v_\mu v_\alpha g_{\beta \nu}  +   v_\nu v_\beta g_{\mu \alpha} \right] 
         + z^{(4)}_4  \left[ v_\mu v_\beta g_{\alpha \nu}  + v_\nu v_\alpha g_{\beta \mu} \right]   \nonumber \\ 
&&    +  w^{(4)} v_\mu v_\alpha v_\beta v_\nu 
 \label{four}
\end{eqnarray} 
Each of the coefficients $a^{(0)} \cdots w^{(4)}$ corresponds to a linearly independent operator, and RPI will imply relations 
between the coefficients $a^{(0)} \cdots w^{(4)}$.

The corresponding total rate is then obtained via the optical theorem by taking a forward matrix element of $R$ with the 
initial state $ | H(p_H) \rangle $
\begin{equation}
2 m_H \Gamma = \langle R \rangle \equiv  \langle H(p_H) | R | H(p_H) \rangle  
\end{equation} 

\subsubsection{The leading order term} 
Applying the RP transformation to the leading term gives
\begin{equation} 
  \delta_{\rm RP}  c^{(0)} (v) = 0 
\end{equation} 
which leads to the RPI result for the leading term 
\begin{equation} \label{R0}
\Gamma = \frac{1}{2 m_H} \langle R \rangle = a^{(0)}   \frac{1}{2 m_H} \langle \phi^\dagger_v \phi_v   \rangle   \, , 
\end{equation} 
in terms of a single matrix element. As we show in the appendix, it is given by 
$$
\langle \phi^\dagger_v \phi_v   \rangle   = 2 m_H \mu_3 = 
2 m_H  \left( 1-    \frac{\mu_\pi^2}{2m^2} \right) \ ,
$$ 
where the last relation is exact to any order in $1/m$ for our definition of $\mu_\pi^2$. 

Before continuing to higher orders, we note that the above result in fact depends 
on $m$ in a nontrivial way. The parameter $\mu_3$ as well as the parameter $\mu_\pi$ 
both depend on $m$; however, in the limit $m \to \infty$ we obtain $\mu_3 = 1$ as expected. 
As we shall see below, the higher-order terms are such that the result is RPI, which becomes manifest by expressing the leading order result in terms of operators 
and matrix elements in full QCD, i.e. with no reference to the arbitrary velocity vector $v$. 
We claim, that this constitutes an improvement of the HQE, since the corresponding 
higher-order terms that will appear in the HQE are now implicitly re-summed in the parameter $\mu_3$.    
 We shall return to this when we have discussed the higher orders of the HQE.

\subsubsection{First and Second Order Terms} 
The first order terms already give insights into the structure of the HQE. 
Applying the RP transformation (\ref{RPI1}) to  $c^{(1)}$ we get a relation between the first and second order terms
\begin{equation}
 \delta_{\rm RP}  c^{(1)}_\mu  (v) =  a^{(1)} \delta v_\mu = m \delta v^\alpha  \, (c^{(2)}_{\alpha \mu} + c^{(2)}_{\mu \alpha } ) 
 = 2 m  \delta v_\mu a^{(2)}  \ ,
 \end{equation} 
which implies $a^{(1)} = 2 m \, a^{(2)}$, while the coefficient $b^{(2)}$ remains unconstrained.  
Inserting this relation into (\ref{opes1}) yields 
\begin{equation}
a^{(1)}   \, \phi^\dagger_v (ivD) \phi_v + a^{(1)}   \,  \frac{1}{2m}  
 \phi^\dagger_v (iD)^2  \phi_v   =   a^{(1)} \, \phi^\dagger_v \left( (ivD) + \frac{1}{2m} (iD)^2 \right)  \phi_v  \ .
\end{equation} 
Note that this particular combination is invariant under reparametrization: 
\begin{equation} \label{RPIivD} 
\delta_{\rm RP}  \left( (ivD) + \frac{1}{2m} (iD)^2 \right)  = 0  \, ; 
\end{equation}  
furthermore, its contribution vanishes when acting on the field $\phi_v$ by the equation of motion.  

In the case at hand it means that we may drop the $(ivD)$ terms as soon as this operator acts directly on the 
field $\phi_v$, since RPI ensures that at the next order a corresponding term with the proper coefficient will 
appear, which will combine with this term to an {\em exactly} vanishing result. Below we show explicitly that this cancellation also appears for higher-order terms. 
 
%
 
In addition,  we re-derive the well-known result that there is no term of linear order in $1/m$ in the HQE; 
this holds true also for the higher-order
terms hidden in $\mu_3$, since according to (\ref{mu3}) the first correction to $\mu_3$ is ${\cal O}(1/m^2)$.
 
\subsubsection{Second and Third Order Terms}
At second order, we obtain
 \begin{eqnarray}  \label{RPxi} 
 \delta_{\rm RP}   c^{(2)}_{\mu \nu} (v) &=& b^{(2)} (\delta v_\mu \, v_\nu + v_\mu \, \delta v_\nu ) \\  \nonumber 
&=& m \delta v^\alpha \left(  c^{(3)}_{\mu  \nu \alpha} + c^{(3)}_{\mu  \alpha  \nu}  + c^{(3)}_{ \alpha  \mu  \nu}  \right)  \\ 
&=& m (x_1^{(3)} + 2 x_2^{(3)} )  [ \delta v_\mu \, v_\nu 
+  \delta v_\nu \, v_\mu ]  \nonumber
\end{eqnarray} 
which implies 
\begin{equation}  \label{RP23} 
m (x_1^{(3)} + 2 x_2^{(3)} ) = b^{(2)} \ .
\end{equation} 
The parameterization in (\ref{four})  of $c^{(3)}$ into the various tensor structures corresponds to a choice of the operator basis, 
such that 
\begin{equation} \label{R3} 
R^{(3)} = x_1^{(3)}  \phi^\dagger_v (iD^\mu) (ivD)  (iD_\mu)  \phi_v  + x_2^{(3)}  \phi^\dagger_v \left\{ (iD)^2 \, , \,  (ivD)  \right\}   \phi_v 
+ x_3^{(3)}  \phi^\dagger_v (ivD)^3 \phi_v  \ .
\end{equation} 
We may solve (\ref{RP23}) for  $x_2^{(3)}$ and insert this into (\ref{R3}) to obtain 
\begin{eqnarray} 
R^{(3)} &=& 
\frac{x_1^{(3)}}{2}  \phi^\dagger_v \left[ (iD_\mu) \, , \, \left[(ivD) \, , \, (iD^\mu) \right] \right]   \phi_v 
+ \frac{b}{2m}   \phi^\dagger_v \left\{ (iD)^2 \, , \,  (ivD)  \right\}   \phi_v + x_3^{(3)}  \phi^\dagger_v (ivD)^3 \phi_v \nonumber \ .
\end{eqnarray}
This relation suggests a change of the operator basis. The first operator generates the well-known Darwin term $\rho_D$ 
and does not relate back to the lower orders in $1/m$. 
RPI fixes the coefficient of the second operator; as discussed in the last section this term combines with the terms of the second order into    
\begin{eqnarray} \label{Comp2} 
&& b^{(2)} \left( \phi^\dagger_v  (ivD)^2 \phi_v + \frac{1}{2m} \phi^\dagger_v \left\{ (ivD) \, , \, (iD)^2 \right\}  \phi_v \right) 
\\ \nonumber && \qquad = b^{(2)} \,  \phi^\dagger_v  \left( (ivD) + \frac{1}{2m} (iD)^2 \right)^2 \phi_v  + {\cal O}(1/m^2) \ ,
\end{eqnarray} 
where the higher-order term is also generated properly, as we show below. The coefficient $c^{(4)}$ of this term can also be related directly to $c^{(2)}$ via the second-order transformation 
$ (\delta_{\rm RP})^2$.


Thus we find that there is no new term generated at order $1/m^2$. The expected $1/m^2$ kinetic energy parameter $\mu_\pi^2$ is contained in the leading order term $\mu_3$. It is well known, that RPI relates the coefficients of the leading term with the one of $\mu_\pi^2$; 
here we suggest to consider this as the RPI completion of the leading-order, now written in terms of $\mu_3$. As we shall see 
below, $\mu_3$ will absorb terms at higher orders in the same way as $\mu_\pi$.   

At order $1/m^3$, we find only a single new term which generates the Darwin term 
with a coefficient that is not constrained 
by RPI; likewise, the coefficient $x_3^{(3)}$ remains unconstrained. However, we shall see below that RPI will again ensure that 
this term gets completed such that the equation of motion can be applied to make it vanish.

\subsubsection{Third  and Fourth Order Terms}
Applying the RPI relation (\ref{RPI1}) to $c^{(3)}$ yields the relation 
\begin{eqnarray} 
\delta_{\rm RP}  c^{(3)}_{\mu \alpha \nu} (v) &=&   
x_1^{(3)}  \delta v_\alpha g_{\mu \nu} + x_2^{(3)}  \left[ \delta v_\nu g_{\mu \alpha}  +  \delta v_\mu g_{\nu \alpha} \right]  
+ x_3^{(3)} \left[ \delta v_\mu \, v_\alpha v_\nu + \delta v_\alpha \, v_\mu v_\nu + \delta v_\nu \, v_\alpha v_\mu  \right]   \nonumber \\ 
 \nonumber 
&=& 2 m (y_1^{(4)} + y_3^{(4)}) \delta v_\alpha  g_{\mu \nu} + 
m (2 y_2^{(4)} + y_1^{(4)} + y_3^{(4)}) \left[ \delta v_\mu g_{\alpha \nu} +  \delta v_\nu g_{\alpha \mu}   \right]  \nonumber \\  
&& + 2 m (z_2^{(4)} + z_4^{(4)})  \delta v_\alpha  \, v_\mu v_\nu + m (2 z_3^{(4)} + z_1^{(4)} 
  + z_4^{(4)}) v_\alpha \left[  \delta v_\mu \, v_\nu + \delta v_\nu \, v_\mu \right]   \, . 
\end{eqnarray} 
Comparing the different tensor structures we obtain the relations 
\begin{eqnarray} \label{RP34}
x_1^{(3)} &=& 2m  \left( y_1^{(4)} + y_3^{(4)} \right)  \ , \\  \label{RP34a}
x_2^{(3)} &=& \frac{b^{(2)}}{2m}  - \frac{x_1^{(3)}}{2} = m \left( 2 y_2^{(4)} + y_1^{(4)} + y_3^{(4)} \right) \ , \\ 
x_3^{(3)} &=& 2 m \left( z_2^{(4)} + z_4^{(4)} \right) = m \left( 2 z_3^{(4)} + z_1^{(4)} + z_4^{(4)} \right)   \ . \vphantom{\frac{1}{1}} 
\label{RP341}
\end{eqnarray} 
There are two contributions to $R^{(4)}$ which are given by 
\begin{eqnarray}
R_1^{(4)} &=& y_1^{(1)} O_1^{(4)} +  y_2^{(1)} O_2^{(4)} + y_3^{(1)} O_3^{(4)}   \\ 
R_2^{(4)} &=& z_1^{(4)} P_1^{(4)} + z_2^{(4)} P_2^{(4)} + z_3^{(4)} P_3^{(4)} + z_4^{(4)} P_4^{(4)} 
\end{eqnarray} 
with the basis operators 
\begin{eqnarray*}
O_1^{(4)} &=& y_1^{(1)} \phi_v^\dagger (iD_\mu) (iD)^2 (iD^\mu) \phi_v  \\ 
O_2^{(4)} &=&  \phi_v^\dagger ((iD)^2)^2  \phi_v   \\ 
O_3^{(1)} &=& \phi_v^\dagger (iD_\mu) (iD_\nu) (iD^\mu) (iD^\nu)\phi_v  \\ 
P_1^{(4)}  &=& \phi_v^\dagger(iD_\mu) (ivD)^2 (iD^\mu) \phi_v \\ 
P_2^{(4)}  &=& \phi_v^\dagger(ivD) (iD)^2 (ivD) \phi_v \\
P_3^{(4)}  &=& \phi_v^\dagger \left\{ (ivD)^2 \, , \, (iD)^2 \right\} \phi_v \\ 
P_4^{(4)} &=&  \phi_v^\dagger \left[ (ivD) (iD_\mu) (ivD) (iD^\mu)  + (iD_\mu) (ivD) (iD^\mu) (ivD) \right]   \phi_v 
\end{eqnarray*}  

Solving the relations (\ref{RP34}, \ref{RP34a}) for $y_1^{(4)}$ and $y_2^{(4)}$ and inserting this into $R_1^{(4)}$ yields
\begin{eqnarray}
 R_1^{(4)} &=& 
\frac{b^{(2)}}{4 m^2} O_2^{(4)} 
 + \frac{x_1^{(3)}}{4m}   \phi_v^\dagger \left[ (iD_\mu) \,  , \, \left[ (iD)^2 \, , \, (iD^\mu) \right] \right] \phi_v \nonumber \\ 
 && + \frac{y_3^{(4)}}{2}  \phi_v^\dagger   [ iD_\mu \, , \,  iD_\nu ]  [ iD^\mu \, , \,  iD^\nu ]  \phi_v \ .
\end{eqnarray} 
The first term is the expected completion of the $(ivD)^2$ in \eqref{Comp2}, while the second term is the 
RPI completion of the Darwin term, 
\begin{equation}  
\phi^\dagger_v \left[ (iD_\mu) \, , \, [ (ivD) \, , \,  (iD^\mu) ] \right]  \phi_v   
\to \phi^\dagger_v \left[ (iD_\mu) \, , \, \left[ \left( ivD + \frac{1}{2 m} (iD)^2\right) \, , \,  (iD^\mu) \right] \right]  \phi_v  \, .
\end{equation} 
Finally, only the coefficient of the third term remains unconstrained 
leading to a genuinely new contribution to  
$R_1^{(4)}$.   

For $R_2^{(4)}$,  we solve  the relations (\ref{RP341}) for $z_2^{(4)}$ and $z_3^{(4)}$ and change to a more convenient base. We find  
 \begin{eqnarray}
 R_2^{(4)} &=& \frac{x_3^{(3)}}{2 m}  \left[ P_2^{(4)}   +   P_3^{(4)}  \right]   -z_1^{(4)} \phi_v^\dagger \left[ (ivD) \, , \, (iD_\mu) \right]  \left[ (ivD) \, , \, (iD^\mu) \right]  \phi_v   \\ \nonumber
 && 
 -\frac{ u^{(4)}}{2}    \phi_v^\dagger \Big\{ (ivD) \, , \,  \left[  (iD^\mu) \, , \, \left[ (iD_\mu) \, , \, (ivD) \right] \right] \Big\}   \phi_v  \ ,
 \end{eqnarray} 
 with $u^{(4)} = z_1^{(4)} + z_4^{(4)}$.   The first term is part of the RPI completion of the $(ivD)^3$ term appearing in the third order. The remaining terms are not constrained by RPI, the second term can be interpreted  as the square of the chromo-electric field, while the last term will vanish by the equations of motion.

In summary, we find that at tree-level, the total rate for scalar-quark QCD up to the order $1/m^4$ can be written in terms of four parameters 
only. We define these parameters as
\begin{eqnarray} \label{param1}
&& \langle \phi_v^\dagger \phi_v \rangle = 2 m_H  \, \mu_3  \vphantom{\frac{1}{1}} \\ 
&&\frac{1}{2} \left\langle \phi_v^\dagger   \left[ (iD_\mu) \, , \, \left[ \left( ivD + \frac{1}{2m} (iD)^2 \right)  
       \, , \,  (iD^\mu) \right] \right] \phi_v \right\rangle = 2 m_H \, \rho_D^3  \\ 
&& \langle  \phi_v^\dagger [ iD_\mu \, , \,  iD_\nu ]  [ iD^\mu \, , \,  iD^\nu ]   \phi_v \rangle = 2 m_H \, r^4_{G} \vphantom{\frac{1}{1}} \\
&&   \langle  \phi_v^\dagger \left[ (ivD) \, , \, (iD_\mu) \right]  \left[ (ivD) \, , \, (iD^\mu) \right]  \phi_v \rangle = 2 m_H \, r^4_{E} 
\vphantom{\frac{1}{1}} \label{Esq} 
\label{param4}
\end{eqnarray} 

In particular, we note the absence of operators such as $[(iD)^2]^2$. This can be understood as a consequence 
of Lorentz invariance. The argument becomes 
particularly simple, if we ignore the presence of gluons and evaluate the forward matrix element 
of $R$ between free quark states. This matrix element will be a Lorentz invariant quantity and hence will 
depend on the square of the quark momentum $p$.  Inserting $p = mv + k$ yields $p^2 = m^2 + 2 m (vk) + k^2$, and by 
the equation of motion we find $2m (vk) + k^2 = 0$ ensuring that $p^2 = m^2$. Hence all terms but the leading one 
vanish by the equation of motion, i.e. all operators involving $k^2$ and $(vk)$ will appear only in the particular combination dictated 
by the equation of motion.   

At this point it is also convenient to compare the above formulation with the one where the covariant derivative is split into 
a spatial and a time derivative, according to 
\begin{equation}
i D_\mu = v_\mu (ivD) + D_\mu^\perp  \, . 
\end{equation}  
While this splitting is very useful in different contexts, it is not useful for the present investigation. In fact, rewriting the HQE in terms 
of operators involving $(ivD)$ and $iD^\perp$ will again re-arrange the terms, without giving additional insights. 

\subsubsection{Re-summation and Relation to full QCD: Scalar Toy Model} \label{sec:FullQCD} 
The parameters we found up to order $1/m^4$ depend on the mass of the heavy quark in a nontrivial way and 
imply a re-summation of higher orders of the HQE in such a way that the final result is actually RPI. This fact can be 
made manifest by re-writing the matrix elements in terms of QCD states and operators. 

While the states are already the ones of full QCD, we still need to un-do the phase redefinition of the quark fields. 
For the leading term $\mu_3$ this obvious, since we have 
$$
\langle \phi^\dagger_v \phi_v   \rangle  = 2m \, \langle \phi^\dagger \phi   \rangle \ ,
$$ 
and thus $\mu_3$ is a matrix element defined in full QCD. 

The next term is the Darwin term $\rho_D$, for which we use the relation  
\begin{equation}
e^{- i m vx}  \left( ivD + \frac{1}{2 m} (iD)^2\right) = \frac{1}{2m}  ((iD)^2 - m^2) e^{- i m vx} \ .
\end{equation} 
Since the mass term does not contribute in the commutator, we find 
\begin{equation}
 \phi^\dagger_v \left[ (iD_\mu) \, , \, \left[ \left( ivD + \frac{1}{2 m} (iD)^2\right) \, , \,  (iD^\mu) \right] \right]  \phi_v 
 =  \phi^\dagger  \left[ (iD_\mu) \, , \, \left[  (iD)^2  \, , \,  (iD^\mu) \right] \right]  \phi \ . 
\end{equation}  
In fact, the Darwin term is related to the chromo-electric field $ E \sim [iD_\mu \, , \, ivD ]$ which is a quantity 
defined in a specific frame. RPI ensures that 
\begin{equation}
  \phi^\dagger_v ...  [iD_\mu \, , \, ivD ] ... \phi_v 
  \to    \phi^\dagger_v ...  \left[iD_\mu \, , \, \left( ivD  + \frac{1}{2m} (iD)^2 \right) \right] ... \phi_v , 
\end{equation} 
where the ellipses denote any combination of derivatives or other operators involving the light degrees of freedom. 
Replacing the field $\phi_v$ by $\phi$ yields 
\begin{eqnarray} 
&&   \phi^\dagger_v ...  \left[iD_\mu \, , \, \left( ivD  + \frac{1}{2m} (iD)^2 \right) \right] ... \phi_v =  \\ \nonumber
&&     \phi^\dagger ...  \left[iD_\mu \, , \,   (iD)^2  \right] ... \phi  =   
   \phi^\dagger ... \left\{ iD_\alpha \, , \,  \left[iD_\mu \, , \,   iD^\alpha \right] \right\}  ... \phi \, .
\end{eqnarray} 
Taking a matrix element of this operator with momentum eigenstates shows that this indeed becomes 
the chromo-electric field in the rest frame of the this state. 

In a similar way the remaining terms can be re-expressed in terms of full QCD and become 
\begin{eqnarray} 
&& \langle  \phi_v^\dagger [ iD_\mu \, , \,  iD_\nu ]  [ iD^\mu \, , \,  iD^\nu ]   \phi_v \rangle 
= 2 m  \langle  \phi^\dagger [ iD_\mu \, , \,  iD_\nu ]  [ iD^\mu \, , \,  iD^\nu ]   \phi \rangle \ , \\
&&   \langle  \phi_v^\dagger \left[ (ivD) \, , \, (iD_\mu) \right]  \left[ (ivD) \, , \, (iD^\mu) \right]  \phi_v \rangle   
=  2 m \langle  \phi^\dagger \left[(iD)^2   \, , \, (iD_\mu)\right]  \left[ (iD)^2 \, , \, (iD^\mu) \right]  \phi   \rangle  . 
\end{eqnarray} 

Note that the power counting is now much less obvious, since the power in $1/m$ is no longer simply related to the 
number of derivatives appearing in the operators. Nevertheless, the leading term in a $1/m$ expansion 
of the matrix elements always reproduces the proper static limit, and the higher order terms are arranged such that 
the final result is RPI. 

\subsubsection{Generalization to Arbitrary Orders}
From the above arguments it becomes clear that one may systematically access higher orders by an iterative process. 
Starting from a suitable tensor decomposition of the coefficients $c^{(n)}$ and  $c^{(n+1)}$ one makes use of 
(\ref{RPI1}) to obtain relations between the coefficients of the  tensor decomposition of $c^{(n)}$ and  $c^{(n+1)}$. 
Taking into account the information obtained from lower orders $m$, $m \le n$ one can determine the elements of the 
operator basis which are constrained by RPI and the ones which emerge as new parameters. However, genuinely new 
matrix elements are only the ones where no $(ivD) $ factor appears next to the field $\phi_v$, since such a contribution 
will vanish exactly once it is properly combined with higher orders. 
 
Finally, in order to make the invariance under reparametrization manifest, one always can re-write the operators and covariant derivatives 
in terms of full QCD operators, which are without any reference to the velocity vector. 

\section{Real Quarks} \label{sec:real} 
Taking into account the quark spin does not change the general idea, the discussion become only a bit more tedious. We start with Eq.~(\ref{opes}) in real QCD 
\begin{equation} \label{ope}
R(S)  =  \sum_{n=0}^\infty  C_{\mu_1 \cdots \mu_n}^{(n)} (S)  \otimes \bar{Q}_v (iD_{\mu_1} \cdots iD_{\mu_n})  Q_v  \, . 
\end{equation}  
Here $\otimes$ is a short hand for the Dirac structure:
\begin{eqnarray}
R(S)  &=&  \sum_{n=0}^\infty  \left[  C_{\mu_1 \cdots \mu_n}^{(n)} (S) \right]_{\alpha \beta}    
\bar{Q}_{v, \alpha}  (iD_{\mu_1} \cdots iD_{\mu_n})  Q_{v, \beta}   \\ \nonumber
&=&  \sum_{n=0}^\infty \sum_\Gamma C_{\mu_1 \cdots \mu_n}^{(n,\Gamma)}   
 \bar{Q}_v (iD_{\mu_1} \cdots iD_{\mu_n})  \Gamma Q_v \ , 
\end{eqnarray}   
where the sum over $\Gamma$ runs over the basis of the 16 Dirac matrices 
$1, \gamma_\mu, \sigma_{\mu \nu}, \gamma_5, i \gamma_\mu \gamma_5$ and 
$$
C_{\mu_1 \cdots \mu_n}^{(n,\Gamma)}  =\frac{1}{4} {\rm Tr} [ \Gamma \, C_{\mu_1 \cdots \mu_n}^{(n)} ] \ .
$$

Applying the RP transformation (\ref{RPT1},\ref{RPT2}, \ref{RPT3}), we arrive at the RPI relation   
\begin{eqnarray} \label{RPI11}
\delta_{\rm RP} C_{\mu_1 \cdots \mu_n}^{(n)} &=& 
m \, \delta v^{\alpha}  \left( C_{\alpha \mu_1 \cdots \mu_n}^{(n+1)}  +  C_{ \mu_1 \alpha \mu_2 \cdots \mu_n}^{(n+1)}   
+ \cdots + C_{ \mu_1 \cdots \mu_n \alpha}^{(n+1)} \right)  \qquad n = 0,1,2, ... 
\end{eqnarray}   
The difference with (\ref{RPI1}) is that the coefficients are now Dirac-matrix valued. 

\subsection{Total Rate for Real Quarks}
For the total rate, the coefficients still depend only on the velocity $v$, where now we have to take into account the spinor structure of the coefficients. The first few terms read 
\begin{equation}
R = \bar{Q}_v C^{(0)} (v) Q_v + \bar{Q}_v C^{(1)}_\mu  (v) (i D^\mu)  Q_v + \bar{Q}_v C^{(2)}_{\mu \nu}  (v)  (i D^\mu)  (i D^\nu)  Q_v
+ \cdots
\end{equation}
The coefficients up to $1/m^2$ are  
\begin{eqnarray}
C^{(0)} (v) &=& a_0   + \hat{a}_0  \fmslash{v}    \ , \\
C^{(1)}_\mu (v) &=&    v_\mu \left( a_1 +  \hat{a}_1  \fmslash{v}  \vphantom{ \hat{b}_2 }   \right)  
+  \gamma_\mu \left(  b_1  +  \hat{b}_1    \fmslash{v}  \right)\ ,  \\
C^{(2)}_{\mu \nu} (v) &=&  v_\mu v_\nu \left( a_2 + \hat{a}_2 \fmslash{v}  \vphantom{ \hat{b}_2 }  \right)  
+  g_{\mu\nu} \left( b_2 + \hat{b}_2  \fmslash{v} \right) 
+   (v_\mu \gamma_\nu + v_\nu \gamma_\mu) \left( d_2 + \hat{d}_2 \fmslash{v} \right)   
+ g_2 (-i \sigma_{\mu\nu}) 
\end{eqnarray} 
where the coefficients $a_1 ... g_2$ are only functions of the quark masses and of the strong coupling $\alpha_s$ and we only consider hermitian operators. We have dropped all parity-odd contributions, since we only discuss ground-state mesons. 

\subsubsection{The leading order} 
Employing now relation (\ref{RPI11}) to the leading coefficient we get 
\begin{equation} 
\delta_{\rm RP} C^{(0)} =  \hat{a}_0 \, \,  \delta v^{\alpha}  \gamma_\alpha \stackrel{\mbox{\tiny RPI}}{=} 
m \, \delta v^{\alpha}  \, C_\alpha^{(1)}  = m \, \delta v^{\alpha} 
  \gamma_\alpha \left(  b_1  +  \hat{b}_1    \fmslash{v}  \right)  \ .
\end{equation}   
Comparing the Dirac and the tensor structure, we obtain the relations  
\begin{equation}  
b_1  = \frac{1}{m} \hat{a}_0 \qquad \mbox{and} \qquad \hat{b}_1 = 0 \ , 
\end{equation} 
while $a_1$ and $\hat{a}_1$ remain unconstrained. 
Gathering the leading and the first order term yields
\begin{eqnarray}
R &=& 
(a_0 + \hat{a}_0)  \bar{Q}_v  Q_v + a_1  \bar{Q}_v  (ivD) Q_v  
+ \hat{a}_1  \bar{Q}_v  (ivD) \fmslash{v}  Q_v  \ ,
\end{eqnarray} 
where the two leading coefficients $a_0$ and $\hat{a}_0$ are related by the equation of motion (\ref{eom1}). As we shall see, this feature will also be present in higher orders. 

The leading term is given by the matrix element $\mu_3$ defined in the appendix. Furthermore, 
RPI enforces that the contribution proportional to $\hat{a}_0$ involving $\fmslash{v}$ is related to the 
term with $\gamma_\alpha$ proportional to $ b_1$. As we shall see below, this will eventually allow us to replace $\fmslash{v} \to 1$, i.e. there will be no contribution with a single $\gamma_\alpha$ matrix.  

\subsubsection{First and Second Order terms} 
In the next step we apply (\ref{RPI11}) to the first order term to obtain
\begin{eqnarray}
\delta_{\rm RP} C^{(1)}_\mu &=& \delta v_\mu \left( a_1 +  \hat{a}_1  \fmslash{v}  \vphantom{ \hat{b}_2 }   \right)  
+ \left( \hat{a}_1 v_\mu + \hat{b}_1 \gamma_\mu \right) \delta \fmslash{v}  \\ 
& \stackrel{\mbox{\tiny RPI}}{=} & m \delta v^\alpha \left(C^{(2)}_{\mu \alpha} + C^{(2)}_{\alpha \mu} \right)  \\ \nonumber
&=& m \delta v^\alpha \left[ 2 g_{\mu \alpha}  \left( b_2 + \hat{b}_2  \fmslash{v} \right)  + 2 \gamma_\alpha v_\mu 
  \left( d_2 + \hat{d}_2 \fmslash{v} \right)  \right] 
\end{eqnarray} 
from which we obtain the relations  
\begin{equation}  \label{RPIrel}
b_2 = \frac{1}{2m} a_1  \qquad \hat{b}_2 = \frac{1}{2m} \hat{a}_1 
 \qquad d_2  = \frac{1}{2m} \hat{a}_1  \quad  \hat{d}_2 = 0  \, . 
\end{equation} 

Collecting all the (non-zero) terms of up to order $1/m^2$ we get 
\begin{eqnarray}
R &=& (a_0 + \hat{a}_0)  \bar{Q}_v  Q_v  + a_1  \bar{Q}_v \left(  (ivD) + \frac{1}{2m} (iD)^2 \right) Q_v   \nonumber \\ 
&& + \hat{a}_1  \bar{Q}_v \left\{ \left(  (ivD) + \frac{1}{2m} (iD)^2 \right)  \, , 
      \, \left( \fmslash{v} +  \frac{1}{m} (i \fmslash{D}) \right) \right\}  Q_v     \nonumber \\ 
&& +  \vphantom{\frac{1}{1}}   g_2  \bar{Q}_v (\sigma \cdot G)  Q_v 
+ \cdots  \ , 
\end{eqnarray} 
where the ellipses denote terms of higher order and terms that are total derivatives; the latter do not contribute 
to the relevant forward matrix elements.

Similar to what happened in the leading order, the terms with $\fmslash{v}$ combine with the corresponding terms 
at the next order to yield the equation of motion. Eventually this means that these terms may be lumped into the 
contributions with the unit Dirac matrix. To this end, the Dirac decomposition of the coefficients $C^{(n)}$ can be 
reduced to the terms with $1$ and $\sigma_{\mu \nu}$.

Furthermore,  the equation of motion (\ref{eom2}) now yields  
\begin{equation}\label{eq:eomspin}
\left(  (ivD) + \frac{1}{2m} (iD)^2 \right) Q_v =  - \frac{1}{2m} (\sigma \cdot G) Q_v \ ,
\end{equation}
where the left and the right hand side are both RPI. 
Finally, 
\begin{equation} 
R = \left( a_0 +  \hat{a}_0 \right)  \bar{Q}_v  Q_v + \left( g_2 - \frac{a_1+ \hat{a}_1 }{2m} \right)     
\bar{Q}_v  ( \sigma \cdot G) Q_v +  {\cal O} (1/m^3) \ .
\end{equation} 

This expression is a re-derivation of the known result, that the HQE does not have 
$1/m$ contributions. Furthermore, up to order $1/m^2$ the HQE
contains two 
non-perturbative parameters $\mu_3$ and $\mu_G$ (or equivalently $\mu_\pi$ and $\mu_G$) which we have 
defined in the appendix.  

\subsubsection{Second and Third Order Terms} 
Dropping all terms with $\fmslash{v}$ and 
single $\gamma$ matrices, we only consider 
\begin{eqnarray}
C^{(2)}_{\mu \nu} (v) &=&  v_\mu v_\nu  a_2  +  g_{\mu\nu}  b_2  + g_2  (-i \sigma_{\mu\nu})    \\ 
C^{(3)}_{\mu \alpha \nu}  (v) &=& x_1^{(3)}  v_\alpha g_{\mu \nu} + x_2^{(3)}  \left( v_\nu g_{\mu \alpha}  +  v_\mu g_{\nu \alpha} \right) 
 + x_3^{(3)}  v_\mu v_\alpha v_\nu   \nonumber \\ 
&& + \xi_1^{(3)} v_\alpha (-i \sigma_{\mu \nu}) 
+ \xi_2^{(3)} \left(  v_\nu (-i \sigma_{\mu \alpha})  +   v_\mu (-i \sigma_{\alpha \nu }) \right)  \, . 
\end{eqnarray}  

The spin independent terms (i.e. the ones without a $\sigma$ matrix) yield the same result as for the scalar case. However, 
the first term in $C^{(2)}$  will generate a term with $(ivD)^2$ which will combine in the same way as in the scalar case to the 
RPI combination in \eqref{eq:eomspin}, which now generates a contribution of $1/(4m^2)  (\sigma \cdot G)^2 $. These terms will appear in the fourth order. 

The spin-dependent (denoted by the superscript $\sigma$)  terms yield 
\begin{eqnarray} 
\delta_{\rm RP} C^{(2, \sigma)}_{\mu \nu} = 0 &=& m \delta v^\alpha  \nonumber
\left( C^{(3, \sigma)}_{\mu  \nu \alpha} + C^{(3, \sigma)}_{\mu  \alpha  \nu}  + C^{(3, \sigma)}_{ \alpha  \mu  \nu}  \right)  \\ 
&& =  m \, \xi_1^{(3)} \delta v^\alpha ( \sigma_{\mu \alpha} v_\nu + \sigma_{\alpha \nu } v_\mu ) \ .
\end{eqnarray} 
From this we conclude that   $\xi_1^{(3)}  =0$ and thus we find that in total rates the usual spin-orbit term $\rho_{LS}$ is absent; this has been 
noticed already in previous papers \cite{Mannel:2010wj,Mannel:2017jfk}. 
This is related to the definition of $\mu_G$ in (\ref{muG}) in terms of the full covariant derivative instead of the spatial components only. Using the definition of $\mu_G$ with spatial 
components only yields an expression which is not RPI, rather it is related by reparametrization to $\rho_{LS}$ 
and hence the corresponding 
combination can be treated as a single parameter, i.e. by our definition of $\mu_G$.    

The remaining term in $C^{(3)}$ contains an $(ivD)$ which acts on $Q_v$. This term will 
be completed in a reparametrization-invariant way at higher orders, rendering a fourth-order contribution.


Thus we find at order $1/m^3$ only one ``genuine'' contribution, which will generate the Darwin term $\rho_D$ with the operator
stucture 
$$ 
\bar{Q}_v \left[ (iD_\mu) \, , \, [ (ivD) \, , \,  (iD^\mu) ] \right]  Q_v    \, . 
$$ 

\subsection{Third and Fourth Order Terms} 
At the fourth order we obtain the same results as in the scalar case. For the additional spin-dependent terms, we find
\begin{eqnarray}
C^{(4 \, \sigma g)}_{\mu \alpha \beta \nu} &=&  
      \alpha_1^{(4)} (-i \sigma_{\mu \nu}) g_{\alpha \beta} 
  +  \alpha_2^{(4)} (-i \sigma_{\alpha \beta}) g_{\mu \nu}   \\ \nonumber 
  && + \alpha_3^{(4)}  \left[ (-i \sigma_{\mu \alpha}) g_{ \beta \nu } +  (-i \sigma_{\beta \nu}) g_{\mu \alpha} \right] 
  + \alpha_4^{(4)} \left[ (-i \sigma_{\mu \beta}) g_{ \alpha \nu } + (-i \sigma_{\alpha \nu}) g_{ \mu \beta }  \right] \ , 
\end{eqnarray} 
corresponding to a linear combination of four hermitian operators:
\begin{equation}
 R^{(4,\sigma)}_1 =  \alpha_1^{(4)} S_1^{(4)} +  \alpha_2^{(4)} S_2^{(4)} +  \alpha_3^{(4)} S_3^{(4)} +  \alpha_4^{(4)} S_4^{(4)}  \ ,
\end{equation} 
with 
\begin{eqnarray}
S^{(4)}_1 &=&  \bar{Q}_v (iD_\mu) (iD)^2 (iD_\nu)(-i \sigma^{\mu \nu}) Q_v \ , \\ 
S^{(4)}_2 &=&  \bar{Q}_v (iD_\alpha) (\sigma \cdot G) (iD^\alpha) Q_v   \ , \\
S^{(4)}_3 &=& \bar{Q}_v \left\{ (iD)^2 \, , \,  (\sigma \cdot G)      \right\} Q_v  \ , \\
S^{(4)}_4 &=& \bar{Q}_v \left[ (iD_\mu) (iD_\alpha) (iD^\mu) (iD_\beta)  
      + (iD_\alpha) (iD^\mu) (iD_\beta)  (iD_\mu)  \right]  (-i \sigma^{\alpha \beta}) Q_v \ .
\end{eqnarray} 

The reparametrization (\ref{RPI11}) relates these terms to the spin-dependent ones in $C^{(3)}$:
\begin{eqnarray}
\delta_{\rm RP} C^{(3,\sigma)}_{\mu \alpha \nu} (v) &=& \xi_2^{(3)} \left(  \delta v_\nu \, (-i \sigma_{\mu \alpha})  +   \delta v_\mu \, (-i \sigma_{\alpha \nu }) \right) \\ 
&=& 2 m (\alpha_1^{(4)} + \alpha_4^{(4)}  ) (-i \sigma_{\mu \nu}) \delta v_\alpha + 
m ( 2 \alpha_3^{(4)} + \alpha_2^{(4)} + \alpha_4^{(4)} ) [  \delta v_\nu \, (-i \sigma_{\mu \alpha})  +   \delta v_\mu \, (-i \sigma_{\alpha \nu }) ] 
\nonumber \\ \nonumber 
&& + m  (\alpha_1^{(4)} + \alpha_4^{(4)}  ) \delta v^\beta \, [   (-i \sigma_{\mu \beta}) g_{\alpha \nu} 
+   (-i \sigma_{\beta \nu}) g_{\mu \alpha} ]  Q_v  \ .
\end{eqnarray} 

From this relation we obtain the equations
\begin{eqnarray}
m (2 \alpha_3^{(4)} + \alpha_2^{(4)} + \alpha_4^{(4)} ) &=&  \xi_2^{(3)} \ , \\ 
\alpha_1^{(4)} + \alpha_4^{(4)}   &=& 0 \ .
\end{eqnarray} 

Solving these equations for $\alpha_3^{(4)}$ and inserting this into $R^{(4,\sigma)}_1 $ yields 
\begin{eqnarray}
R^{(4,\sigma)}_1 &=&
\frac{ \xi_2^{(3)}}{2m}  S_3^{(4)} 
 + \alpha_1^{(4)} \left[ S_1^{(4)} + S_2^{(4)} - S_4^{(4)}  + \left( \frac{S_3^{(4)}}{2} - S_2^{(4)}\right)  \right] 
- \frac{\alpha_2^{(4)}}{2} \left[ S_3^{(4)} - 2 S_2^{(4)} \right]   
\end{eqnarray}  
The first term is the expected completion of terms appearing in the third order 
$$ 
\bar{Q}_v \left\{ (ivD) \, , \,  (\sigma \cdot G) \right\} Q_v \to  
\bar{Q}_v \left\{ \left( ivD + \frac{1}{2m} (iD)^2 \right) \, , \,  (\sigma \cdot G) \right\} Q_v \ ,
$$ 
while the remaining terms remain unconstrained and have a 
simple physical interpretation in terms of the operators 
\begin{eqnarray} 
 S_1^{(4)} +  S_2^{(4)} -  S_4^{(4)} &=& 
\bar{Q}_v \left[ (iD_\mu) \, , \,  (iD_\alpha) \right]  \left[ (iD_\beta)  \, , \, (i D^\mu) \right] (-i \sigma^{\alpha \beta})  Q_v \ , \\ 
 S_3^{(4)} - 2  S_2^{(4)}  &=&  
\bar{Q}_v \left[ (iD_\mu) \, , \, \left[  (iD_\mu) \, , \, (\sigma \cdot G) \right] \right] Q_v  \ .
\end{eqnarray} 
The matrix element of the first operator is related to $\sigma \cdot (G \times G)$, while the second operator is related to ${\cal D}^2 (\sigma \cdot G)$, where ${\cal D}$ is the covariant derivative in the adjoint representation, acting on $G$. 

The second contribution can be parametrized as 
\begin{eqnarray}
C^{(4 \, \sigma vv)}_{\mu \alpha \beta \nu} &=&  
      \beta_1^{(4)} (-i \sigma_{\mu \nu}) v_\alpha v_\beta 
  +  \beta_2^{(4)} (-i \sigma_{\alpha \beta}) v_\mu v_\nu   \\ \nonumber && 
    +  \beta_3^{(4)} \left[  (-i \sigma_{\mu \alpha}) v_\nu v_\beta  
    +  (-i \sigma_{\nu \beta}) v_\mu v_\alpha \right]  
            +  \beta_4^{(4)} \left[ (-i \sigma_{\nu \alpha}) v_\mu v_\beta +  (-i \sigma_{\mu \beta}) v_\nu v_\alpha   \right] \ ,
\end{eqnarray}  
corresponding to the linear combination of operators 
\begin{equation}
R_2^{(4, \sigma)} =   \beta_1^{(4)} U_1^{(4)} +  \beta_2^{(4)} U_2^{(4)} +  \beta_3^{(4)} U_3^{(4)} +  \beta_4^{(4)} U_4^{(4)} \ ,
\end{equation} 
with 
\begin{eqnarray}
 U_1^{(4)}  &=& \bar{Q}_v (iD_\mu) (ivD)^2 (iD_\nu)  (-i \sigma^{\mu \nu}) Q_v \ ,  \\ 
 U_2^{(4)}  &=& \bar{Q}_v  (ivD) (\sigma \cdot G) (ivD)   Q_v  \ ,\\ 
 U_3^{(4)}  &=& \bar{Q}_v  \left\{ (ivD)^2 \, , \, (\sigma \cdot G)  \right\}   Q_v  \ , \\ 
 U_4^{(4)} &=& \bar{Q}_v \left[ (ivD ) (iD_\alpha) (ivD) (iD_\beta)  
      + (iD_\alpha) (ivD) (iD_\beta)  (ivD)  \right]   (-i \sigma^{\alpha \beta}) Q_v  \ .
\end{eqnarray} 
Using the reparametrization relation (\ref{RPI11}) we find no terms of this form in $\delta_{\rm RP} C^{(3)} $ and thus
\begin{equation}
0 = m (  \beta_1^{(4)} +   \beta_4^{(4)} ) 
  \delta v^\beta \, \left[ v_\mu v_\alpha (-i \sigma_{\nu \beta}) +  v_\nu v_\alpha (-i \sigma_{\beta \nu})
\right] \ ,  
\end{equation} 
from which we conclude 
\begin{equation}
  \beta_1^{(4)} = -   \beta_4^{(4)}  \ ,
\end{equation} 
while all other operator coefficients remain unconstrained.   Inserting this into $R_2^{(4, \sigma)}$, we write  
\begin{eqnarray}
R_2^{(4, \sigma)}
&=&  (\beta_2^{(4)} -  \beta_2^{(4)} ) U_2^{(4)} +  \beta_3^{(4)} U_3^{(4)}  + \beta_1^{(4)} \left[ U_1^{(4)}  + U_2^{(4)}  -   U_4^{(4)} \right]  \ .
\end{eqnarray}  
The operators $U_2^{(4)}$ and $U_3^{(4)}$ have $(ivD)$ factors acting directly on $Q_v$ and   
thus will contribute only to higher orders,  
while the only non-vanishing contribution at order $1/m^4$  is  
\begin{equation}
U_1^{(4)} + U_2^{(4)}  -   U_4^{(4)}  = \bar{Q}_v \left[ D_\mu   \, , \, ivD  \right]  \left[ ivD  \, , \, iD_\nu \right]  (-i \sigma^{\mu \nu}) Q_v   \, . 
\end{equation} 
The matrix element of this operator corresponds to the product $\sigma \cdot (\vec{E} \times \vec{E})$ where  $\vec{E}$ is the 
chromo-electric field.

\subsection{Resummation and Relation to full QCD: Real Quarks} 
Up to order $1/m^4$, we find in total eight independent parameters at tree level, defined by the matrix elements 
\begin{eqnarray}\label{eq:MEs}
&& \langle \bar{Q}_v Q_v \rangle = 2 m_H \mu_3 \vphantom{\frac{1}{1} } \ ,\\
&& \langle \bar{Q}_v  (i D_\alpha) (i D_\beta) (-i \sigma^{\alpha \beta} ) Q_v \rangle =  2m_H d_H \mu_G^2  \vphantom{\frac{1}{1} } \ , \\ 
&& \frac{1}{2}\langle \bar{Q}_v \left[ (iD_\mu) \, , \,   \left[ \left( i vD + \frac{1}{2m} (iD)^2 \right)  \, , \, (i D^\mu) \right] \right]   Q_v \rangle 
= 2 m_H \rho_D^3 \ ,\\ 
&& \langle \bar{Q}_v \left[ (iD_\mu) \, , \,  (iD_\nu) \right]  \left[ (iD^\mu)  \, , \, (i D^\nu) \right]   Q_v \rangle 
= 2 m_H r_{G}^4  \vphantom{\frac{1}{1} }\ ,  \\ 
\label{eq:rE4}&& \langle \bar{Q}_v \left[ (ivD ) \, , \,  (iD_\mu) \right]  \left[ (ivD)  \, , \, (i D^\mu) \right]   Q_v \rangle 
= 2 m_H r_{E}^4  \vphantom{\frac{1}{1} } \ , \\
&& \langle \bar{Q}_v \left[ (iD_\mu) \, , \,  (iD_\alpha) \right]  \left[ (iD^\mu)  \, , \, (i D_\beta) \right]   (-i \sigma^{\alpha \beta})  Q_v \rangle 
= 2 m_H d_H s_{B}^4  \vphantom{\frac{1}{1} }\ ,  \\ 
&& \langle \bar{Q}_v \left[ (ivD) \, , \,  (iD_\alpha) \right]  \left[ (ivD)  \, , \, (i D_\beta) \right]   (-i \sigma^{\alpha \beta})  Q_v \rangle 
= 2 m_H d_H s_{E}^4  \vphantom{\frac{1}{1} }\ ,  \\ 
&& \langle \bar{Q}_v  \left[ iD_\mu \, , \, \left[ iD^\mu \, , \,  \left[ iD_\alpha \, , \, iD_\beta \right] \right] \right]  (-i \sigma^{\alpha \beta}) Q_v 
\rangle = 2 m_H d_H s_{qB}^4  \vphantom{\frac{1}{1} } \ ,
\end{eqnarray} 
where $d_H= 1$ for mesons and $d_H=0$ for baryons. 
We note that these operators contain higher orders of $1/m$ in such a way that the result is RPI to all orders. The proper power counting 
can still be performed, since the contributions appearing at order $1/m^n$ do not  contain any pieces of powers $1/m^k$ with  $k \le n$. 
Thus the result is correct to order $1/m^n$, but is fully RPI.  

We have chosen these operators in such a way that they have a clear physical interpretation. We have 
\begin{eqnarray}
 &&\mu_G^2 \sim \langle \bar{Q}_v (\vec{\sigma} \cdot \vec{B}) Q_v \rangle \\
 && \rho_D^3 \sim   \langle \bar{Q}_v ({\rm Div} \vec{E}) Q_v \rangle  \\ 
 && r_{G}^4  \sim \langle \bar{Q}_v (\vec{E}^2- \vec{B}^2) Q_v \rangle \\
 && r_{E}^4  \sim \langle \bar{Q}_v \vec{E}^2 Q_v \rangle \\ 
 && s_{B}^4  \sim \langle \bar{Q}_v (\vec{B} \times \vec{B}) \cdot \vec{\sigma}  Q_v \rangle \\ 
 && s_{E}^4  \sim \langle \bar{Q}_v (\vec{E} \times \vec{E}) \cdot \vec{\sigma}  Q_v \rangle \\ 
 && s_{qB}^4  \sim \langle \bar{Q}_v  ( \Box \, \vec{\sigma} \cdot \vec{B} )   Q_v \rangle 
\end{eqnarray} 
We note that all these operators involve at least one gluon field; in the formal limit $g_s \to 0$ all higher dimensional operators 
vanish and only the leading $\bar{Q}_v Q_v$ remains. 

Comparing our results with those in e.g. \cite{Mannel:2010wj} we notice that the RPI approach yields a smaller number of parameters. 
This is due to the fact that reparametrization strictly links coefficients of some of the parameters listed in  \cite{Mannel:2010wj} and 
hence these parameters are not independent. In the RPI approach advertised here these terms are combined in a single parameter. 

Finally, to make RPI manifest, we may as well express these operators in terms of full QCD operators. As we have shown explicitly for the case of the Darwin term, any appearance of (ivD) will become completed by a higher-order term in an RPI fashion. In terms of full QCD fields and the 
corresponding derivatives this means that
$$
ivD \, Q_v \to \frac{1}{2m} ((iD)^2 - m^2) Q
$$ 
and hence we can write our operators and matrix elements as   
\begin{eqnarray}
&& \langle \bar{Q} Q \rangle = 2 m_H \mu_3 \vphantom{\frac{1}{1} } \ , \\
&& \langle \bar{Q}  (i D_\alpha) (i D_\beta) (-i \sigma^{\alpha \beta} ) Q \rangle =  2m_H d_H \mu_G^2  \vphantom{\frac{1}{1} } \ , \\ 
&& \frac{1}{4m} \langle \bar{Q} \left[ (iD_\mu) \, , \,   \left[ (iD)^2  \, , \, (i D^\mu) \right] \right]   Q \rangle 
= 2 m_H \rho_D^3 \ , \\ 
&& \langle \bar{Q} \left[ (iD_\mu) \, , \,  (iD_\nu) \right]  \left[ (iD^\mu)  \, , \, (i D^\nu) \right]   Q \rangle 
= 2 m_H   r_{G}^4  \vphantom{\frac{1}{1} } \ , \\ 
&& \frac{1}{4m^2}  \langle \bar{Q}  \left[ (iD)^2 \, , \,  (iD_\mu) \right]  \left[ (iD)^2  \, , \, (i D^\mu) \right]   Q \rangle 
= 2 m_H r_{E}^4  \vphantom{\frac{1}{1} }\ ,   \\
&& \langle \bar{Q} \left[ (iD_\mu) \, , \,  (iD_\alpha) \right]  \left[ (iD^\mu)  \, , \, (i D_\beta) \right]   (-i \sigma^{\alpha \beta})  Q \rangle 
= 2 m_H d_H s_{B}^4  \vphantom{\frac{1}{1} } \ , \\ 
&&  \frac{1}{4m^2}  \langle \bar{Q} \left[ (iD)^2 \, , \,  (iD_\alpha) \right]  \left[ (iD)^2  \, , \, (i D_\beta) \right]   (-i \sigma^{\alpha \beta})  Q \rangle 
= 2 m_H d_H s_{E}^4  \vphantom{\frac{1}{1} } \ , \\ 
&& \langle \bar{Q}  \left[ iD_\mu \, , \, \left[ iD^\mu \, , \,  \left[ iD_\alpha \, , \, iD_\beta \right] \right] \right]  (-i \sigma^{\alpha \beta}) Q 
\rangle = 2 m_H  d_H s_{qB}^4  \vphantom{\frac{1}{1} } \ .
\end{eqnarray} 
Note that the power counting becomes now more complicated, since the dimension of the operator (i.e.\ the number of derivatives in the 
operator) no longer corresponds to the order in the $1/m$ expansion.

\section{Beyond tree Level} 
Most of the relations derived in this paper also hold beyond tree level, since RPI must hold 
also beyond tree level.  However, the OPE  (\ref{ope})  must be generalized to include all possible 
operators with the relevant dimension at each order. These operators are built from quark fields 
(light and heavy) and gluon fields as well as from derivatives acting on these fields. We define that all 
light quark and gluon fields as well as the derivatives acting on these fields are invariant under 
reparametrization and thus the behavior of any operator under reparametrization is defined.  
Since the total sum of the OPE is again RPI, the generalization of (\ref{RPI11}) is obvious. 

The operators which appear up to order $1/m^4$ have been written down e.g. in \cite{Mannel:2010wj} 
and their RG mixing has been discussed. Up to order $1/m^2$ the full OPE does not have any additional 
operators beyond the ones we have defined at tree level, which means that our conclusions remain true 
to all orders in $\alpha_s$. At order $1/m^3$ new operators appear, which  are four quark operator of the form 
\begin{equation}\label{4QO}
 ( \bar{Q}_v \Gamma Q_v  ) (\bar{q} \Gamma q )   \;\; \text{and} \;\; ( \bar{Q}_v \Gamma T^a Q_v  ) (\bar{q} T^a \Gamma q )  
\end{equation}  
where $T^a$ are the generators of color $SU(3)$ and $\Gamma$ is a ($v$ independent) Dirac matrix. 
These operators are RPI, therefore their coefficients may not be related by RPI to any other coefficient. Likewise, 
one can directly write these operators in full QCD by simply dropping the subscript $v$ in (\ref{4QO}).   

In some applications it may happen, that the four-quark operators depend on $v$ by a $v$ dependence of 
$\Gamma$ (e.g.\ $\Gamma = \fmslash{v}$). In such a case   (\ref{RPI11})  will relate this operator to a 
term appearing at higher order, a relation that holds at any order in $\alpha_s$. 

Finally, we have to consider the radiative corrections to the color-octet operators. We consider as an illustration $r_E^4$ in (\ref{eq:rE4}).  Writing the chromo-electric field in its components 
\begin{equation}
E_\mu^a T^a = [(ivD) \, , \, (iD_\mu) ]   \ ,
\end{equation} 
we see that $r_E^4$ only contains the (tree level) combination 
\begin{equation}
[(ivD) \, , \, (iD_\mu) ]  [ (ivD)  \, , \, (iD^\mu) ]  = - E_\mu^a   E^{\mu , b}  \,\,    \frac{1}{2}\left\{T^a,  T^b\right\}\  ,
\end{equation} 
where\begin{equation}
 \left\{ T^a,  T^b\right\}  =  d^{abc} T^c + \frac{1}{3} \delta^{ab}  
\end{equation}
and $d^{abc} =   d^{bac}$ the $SU(3)$ symmetric symbol. 
It has been shown that these two contributions receive different radiative corrections \cite{Kobach:2017xkw, Gun17}. Therefore, beyond tree level, $r_E^4$ will generate two independent parameters. 

\section{Alternative Normalization} 
The normalization of the leading term $\mu_3$ is derived from the conservation of the 
heavy flavour current in full QCD (see Appendix 
A).  However, it is worthwhile to point out that there is also an alternative normalization possible which will relate $\bar{Q} Q$ to the hadron mass. 

We start from the energy momentum $\Theta_{\mu \nu}$  tensor of QCD, including the heavy quark. Energy and momentum 
conservation implies 
\begin{equation}
\partial^\mu  \Theta_{\mu \nu} = 0 \ ,
\end{equation} 
which gives the normalization 
\begin{equation}
\langle H(p) |  \Theta_{\mu \nu}  | H(p) \rangle = \langle    \Theta_{\mu \nu}  \rangle = 2 p_\mu p_\nu \ .
\end{equation} 

Using the expression from QCD for the energy momentum tensor and taking the trace, we get 
\begin{equation}
\Theta^\mu_{\,\, \mu}  = m \, \bar{Q} Q + \frac{\beta(\alpha_s)}{4 \pi} G_{\mu \nu}^a G^{\mu \nu, \, a} 
\end{equation} 
where we have added the well-known contribution of the trace anomaly.  Taking the forward matrix element gives 
\begin{equation} 
 \langle    \Theta^\mu_{\,\, \mu}  \rangle = 2 m_H^2 =  m \langle \bar{Q} Q \rangle 
   +  \frac{\beta(\alpha_s)}{4 \pi}  \langle G_{\mu \nu}^a G^{\mu \nu, \, a} \rangle \ ,
\end{equation}  
and hence we obtain
\begin{equation}
m \, \mu_3 =  m_H   - \frac{1}{2 m_H}  \frac{\beta(\alpha_s)}{4 \pi}  \langle G_{\mu \nu}^a G^{\mu \nu, \, a} \rangle \ .
\end{equation}  

We note that inserting (\ref{mu3Def}) yields an exact expression for the hadron mass
 \begin{equation} 
m_H = m  -  \frac{1}{2m} (\mu_\pi^2 - \mu_G^2) + \frac{1}{2 m_H}  \frac{\beta(\alpha_s)}{4 \pi}  \langle G_{\mu \nu}^a G^{\mu \nu, \, a} \rangle \ .
\end{equation}
This can be compared to the  $1/m$ expansion for the pseudoscalar meson ground state:
\begin{equation} 
m_H = m + \bar{\Lambda} + \frac{1}{2m} (\lim_{m \to \infty}  \mu_\pi^2 - \lim_{m \to \infty}  \mu_G^2) +  {\cal O} (1/m^3) \ ,
\end{equation} 
from which we conclude that 
\begin{equation}
 \lim_{m \to \infty}  \frac{1}{2 m_H}  \frac{\beta(\alpha_s)}{4 \pi}  \langle G_{\mu \nu}^a G^{\mu \nu, \, a} \rangle = \bar{\Lambda} = \lim_{m \to \infty} (m_H - m)  \ ,
\end{equation} 
leading to 
\begin{equation} 
 \lim_{m \to \infty}  \mu_3 = 1
\end{equation} 
as expected. 

However, we may use the RPI formulation to write for a weak decay of a heavy hadron $H$ the relation 
\begin{equation}
\Gamma \propto G_F^2 m^5 \mu_3 =  G_F^2 m^4 \left(m_H  - \frac{1}{2 m_H}  \frac{\beta(\alpha_s)}{4 \pi}  \langle G_{\mu \nu}^a G^{\mu \nu, \, a} \rangle \right) 
\end{equation} 
corresponding to the leading order result in the reparametrization-invariant formulation.  


\section{Conclusion} 

We have made use of the fact that RPI relates different orders in the HQE to perform partial re-summations. Computing up to 
a specific order in $1/m$, we combine terms of higher order in such a way that the result becomes RPI. This can be made manifest by writing the resulting matrix elements for the non-perturbative input as matrix elements of 
operators and states defined in full QCD, which do not have any reference to the velocity vector needed 
to set up the HQE. 
  
Clearly the RPI improved results calculated to a certain order contain arbitrarily high orders in $1/m$; however, they are still 
correct only up to the order one has actually calculated, since at each order new terms appear, which are not related by reparametrization to terms 
appearing at lower order. In turn, our approach re-sums all the terms which may be related back to lower-order 
terms, and thus we expect an improvement. 

In this paper, we have confined our discussion to total rates. In this case, a side effect of RPI is that the number of 
independent parameters in the HQE is reduced compared to earlier analyses. While the relations implied by RPI have been found 
some time ago, this has never been used to explicitly reduce the number of independent parameters at $\mathcal{O}(1/m^4)$.

The ``genuine'' terms at higher orders, which do not relate back to lower orders by reparametrization, 
are all due to the presence of gluons or additional light quarks. 
The matching calculation to compute the OPE coefficients  for the total rate is conveniently done using free quark and gluon 
states. In our approach the leading operator $\bar{Q} Q$ is the only contribution which appears in the matching using only the two 
quarks and no gluons; all higher-order terms require at least one gluon in the matching calculation. Therefore,  all the matrix 
elements which have a zero-gluon matrix element are contained in the matrix element of  $\bar{Q} Q$.    

The relations obtained from RPI have been formulated for the general case, i.e. also for differential rates. However, for differential rates 
the RPI relation imply differential equations for the coefficients of the HQE; a detailed analysis for the differential case 
is beyond the scope of this work and will be exploited in future work.

\section*{Acknowledgements} 
We thank Alexei Pivovarov and Gil Paz for discussions related to this subject. This work 
was supported by DFG through the Research Unit FOR 1873 ''Quark Flavour Physics and Effective Field Theories''  

\appendix
\section{Matrix Elements} 

\subsection{Normalization for the Scalar Case}  
Here we collect the expression for the relevant forward matrix elements for the 
scalar case. The leading matrix element is 
\begin{equation}
\langle \phi^\dagger_v \phi_v   \rangle = \langle H(p) |  \phi^\dagger_v \phi_v | H(p) \rangle =  2 m \langle \phi^\dagger \phi   \rangle
\end{equation} 
where $| H(p) \rangle$ is the hadronic state of full (scalar) QCD.  We note that we may re-insert the full QCD operators and define 
\begin{equation}
 \langle \phi^\dagger \phi   \rangle = 4 m_H^2 \mu_3
\end{equation}  
with a hadronic parameter $\mu_3$.  

By a similar argument as for real quarks we can show that $\mu_3 = 1$ up to terms of
order $1/m^2$: We note that the equation of motion (\ref{SG})  for $\phi$ has a conserved current  of the form
\begin{equation} 
J_\mu = \phi^\dagger (i \stackrel{\leftrightarrow}{D}_\mu)  \phi = -i \, \left( (D_\mu \phi)^\dagger \phi - \phi^\dagger (D_\mu \phi)\right) \ .
\end{equation} 
Inserting the rescaled field we get (replacing $\phi$ by $\phi_v$) 
$$ 
J_\mu =  v_\mu \phi_v^\dagger \phi_v +  \frac{1}{2m}\phi_v^\dagger  i \stackrel{\leftrightarrow}{D}_\mu  \phi_v 
$$
Taking the forward matrix element of this operator yields
\begin{equation} 
\langle J_\mu \rangle  = 2 p_\mu  =  2 m_H v_\mu = 
 v_\mu \langle \phi_v^\dagger \phi_v  \rangle + \frac{1}{m} \langle \phi_v^\dagger  i D_\mu  \phi_v \rangle 
\end{equation} 
where we have made a choice for the velocity $v$ to be $v = p_H / m_H$. Contracting with $v^\mu$ we get 
\begin{equation}  \label{norm}
2 m_H
= \langle \phi_v^\dagger \phi_v  \rangle + \frac{1}{m} \langle \phi_v^\dagger  (ivD)   \phi_v \rangle 
=  \langle \phi_v^\dagger \phi_v  \rangle - \frac{1}{2m^2} \langle \phi_v^\dagger  (iD)^2   \phi_v \rangle 
\end{equation} 
Defining the kinetic- energy parameter $\mu_\pi$ as 
\begin{equation}
\langle \phi_v^\dagger  (iD)^2   \phi_v \rangle = - 2 m_H \mu_\pi^2 
\end{equation} 
we finally get  
\begin{equation} \label{mu3}
 \mu_3 = 1-    \frac{\mu_\pi^2}{2m^2} 
\end{equation}  
As discussed in the text, there is no matrix element at order $1/m^2$ in the total rates, beyond $\mu_\pi^2$ 
which only appears in the expression for $\mu_3$. 

\subsection{Real QCD} 
From the equation of motion (\ref{eom1}), we find for a Dirac matrix $\Gamma$ 
\begin{eqnarray}
&& \bar{Q}_v (iD_{\mu_1}) ... (iD_{\mu_n})  \Gamma Q_v = \bar{Q}_v  (iD_{\mu_1}) ... (iD_{\mu_n}) \Gamma \fmslash{v}   Q_v  
+ \frac{1}{m}
\bar{Q}_v   (iD_{\mu_1}) ... (iD_{\mu_n}) \Gamma (i \fmslash{D}) Q_v    \\  
&& \bar{Q}_v  (iD_{\mu_1}) ... (iD_{\mu_n})   \Gamma Q_v = \bar{Q}_v   (iD_{\mu_1}) ... (iD_{\mu_n}) \fmslash{v} \Gamma Q_v 
+ \frac{1}{m}
\bar{Q}_v  (i \fmslash{D})  (iD_{\mu_1}) ... (iD_{\mu_n}) \Gamma Q_v \nonumber \\  
&& \qquad \qquad \qquad \qquad \qquad \qquad + \mbox{total derivative} 
\end{eqnarray} 
where the total derivative will vanish when a forward matrix element is taken.  

We obtain 
\begin{eqnarray}
&& \langle  \bar{Q}_v  (iD_{\mu_1}) ... (iD_{\mu_n})   \Gamma Q_v \rangle 
= \frac{1}{2} \langle  \bar{Q}_v    (iD_{\mu_1}) ... (iD_{\mu_n}) \left\{ \Gamma \, , \, \fmslash{v} \right\}  Q_v \rangle \nonumber \\ 
&& \qquad \qquad  + \frac{1}{2 m}
\langle \bar{Q}_v  \left\{ (i \fmslash{D}) \, , \,  (iD_{\mu_1}) ... (iD_{\mu_n}) \Gamma \right\} Q_v  \ , \rangle  
\end{eqnarray}

which yields for $\Gamma = 1$ 
\begin{eqnarray} \label{one} 
&& \langle  \bar{Q}_v  (iD_{\mu_1}) ... (iD_{\mu_n})  Q_v \rangle 
= \langle  \bar{Q}_v   (iD_{\mu_1}) ... (iD_{\mu_n})  \fmslash{v}  Q_v \rangle \nonumber \\ 
&& \qquad \qquad  + \frac{1}{2 m}
\langle \bar{Q}_v  \left\{ (i \fmslash{D}) \, , \,  (iD_{\mu_1}) ... (iD_{\mu_n}) \right\} Q_v  \rangle \ , 
\end{eqnarray} 
and for $\Gamma = \gamma_\alpha$ 
\begin{eqnarray} \label{two} 
&& \langle  \bar{Q}_v  (iD_{\mu_1}) ... (iD_{\mu_n})  \gamma_\alpha  Q_v \rangle 
=  v_\alpha \langle  \bar{Q}_v  (iD_{\mu_1}) ... (iD_{\mu_n}) Q_v \rangle \nonumber \\ 
&& \qquad \qquad  + \frac{1}{2 m}
\langle \bar{Q}_v  \left\{ (i \fmslash{D}) \, , \,  \gamma_\alpha (iD_{\mu_1}) ... (iD_{\mu_n})  \right\} Q_v  \rangle \ .
\end{eqnarray} 
These relations show that all the contributions with a single $\gamma_\alpha$ can be dropped.

The leading order matrix element is defined as in the scalar case 
\begin{equation}
\langle \bar{Q}_v Q_v \rangle = 2 m_H \mu_3  \ .
\end{equation} 
Using the relation (\ref{two}) for $n=0$ we get 
\begin{equation}
\langle  \bar{Q}_v  \gamma_\alpha Q_v \rangle  =  v_\alpha \langle  \bar{Q}_v  Q_v \rangle  
 + \frac{1}{m}
\langle \bar{Q}_v   (iD_\alpha)    Q_v  \rangle  \ .
\end{equation} 
Contracting with $v^\alpha$ we obtain 
\begin{equation} \label{norma} 
\langle  \bar{Q}_v \fmslash{v}  Q_v \rangle  =  \langle  \bar{Q}_v  Q_v \rangle  
 + \frac{1}{m} \langle \bar{Q}_v   (i v D)    Q_v  \rangle   =  \langle \bar{Q}  \fmslash{v} Q  \rangle = 2 m_H \ , 
\end{equation} 
where we used the conservation of the $b$ quark current in QCD. 
Furthermore, we may use (\ref{eom2}) to obtain
\begin{equation}
 \langle \bar{Q}_v Q_v \rangle =  2 m_H 
 + \frac{1}{2m^2}   \langle \bar{Q}_v  (i \fmslash{D}) (i \fmslash{D}) Q_v \rangle \ .
\end{equation}  
Finally, using the definitions 
\begin{eqnarray}
\langle \bar{Q}_v  (i D)^2   Q_v \rangle   &=& - 2m_H \mu_\pi^2  \ , \\ 
\langle \bar{Q}_v   (\sigma \cdot G) Q_v \rangle & = & 2m_H d_H  \mu_G^2  \label{muG} \ ,
\end{eqnarray} 
where $d_H=1$ for mesons and $d_H=0$ for baryons, we find 
\begin{equation} \label{mu3Def}
\mu_3 = 1 - \frac{1}{2 m^2} (\mu_\pi^2 - d_H \mu_G^2)  \ .
\end{equation}

\section{Example: Tree level $B \to X_s \gamma$}
As an example, we compute the radiative radiative $b \to s \gamma$ decay at tree level but with higher-order $1/m$ corrections. 
For illustration, we consider only the contribution from the operator $O_7$  
\begin{equation}
\frac{\lambda}{2} \bar{s} \sigma_{\mu\nu} (1+\gamma_5)b F^{\mu\nu} \quad \mbox{with} \quad 
\lambda = \frac{e m}{16 \pi^2} | C_7 (m) V_{ts} V_{tb}^* | \ .
\end{equation}
We find, considering massless $s$ quarks,
\begin{equation}
T = - 2 \lambda^2 \, \bar{b}_v \left[ \sigma_{\mu \alpha} q^\alpha \left( \frac{1}{\fmslash{S} + i \fmslash{D}} \right)  \sigma_{\nu \beta} q^\beta g^{\mu \nu} 
\frac{1}{q^2}  \right] b_v \ , 
\end{equation} 
where $S = p - q$, and $q$  is the photon momentum.  
From the expansion of the $s$ quark propagator, we obtain 
\begin{eqnarray}
  \frac{1}{\fmslash{S} + i \fmslash{D}} &=&  \frac{1}{\fmslash{S}} -  \frac{1}{\fmslash{S}} i \fmslash{D}  \frac{1}{\fmslash{S}} + 
   \frac{1}{\fmslash{S}} i \fmslash{D}  \frac{1}{\fmslash{S}}  i \fmslash{D}  \frac{1}{\fmslash{S}} + \cdots  \\ \nonumber
  &=& \frac{1}{S^2} \fmslash{S} -  \left(\frac{1}{S^2} \right)^2 \fmslash{S} i \fmslash{D}  \fmslash{S} + 
   \left(\frac{1}{S^2} \right)^3 \fmslash{S} i \fmslash{D}  \fmslash{S}  i \fmslash{D}  \fmslash{S}  + \cdots \ .
\end{eqnarray}  
Performing the loop integration and taking the imaginary part yields for the total rate;
\begin{equation} 
\Gamma_{bsg}=  \frac{\lambda^2 m^3}{4 \pi} \left[ \mu_3 - \frac{2}{m^2} \mu_G^2 
  - \frac{10 \rho_D^3}{3m^3} - \frac{1}{3m^4} \left(4 r^4_G +4r^4_E +\frac{1}{4} s^4_{qB} - 2 s^4_E \right)\right]  \ ,
\end{equation}
where we indeed see the expected reduction to independent matrix elements. Note that the contribution of $s^4_B$ is absent in 
this relation which is accidental.

\end{document}